\documentstyle[preprint,revtex]{aps}
\begin{document}
\draft
\hfill\vbox{\hbox{\bf NUHEP-TH-92-21}\hbox{Oct 1992}}\par
\thispagestyle{empty}
\begin{title}
{\bf Associated production of Intermediate Higgs or Z-boson  \\
with $t \bar t$ pair in $\gamma \gamma$ collision}
\end{title}
\author{Kingman~Cheung}
\begin{instit}
Dept. of Physics \& Astronomy, Northwestern University, Evanston,
Illinois 60208, USA\\
\end{instit}
\begin{abstract}
\nonum
\section{Abstract}
Photon-photon linear colliders can be realized by laser back-scattering
technique on the next generation linear $e^+e^-$ colliders.
Here the associated productions of an intermediate mass
Higgs (IMH) or Z-boson with $t \bar t$ pair in $\gamma\gamma$ collisions
are studied.  Since IMH is very unlikely to decay into $t\bar t$ pair,
$t\bar tH$ production is the only direct channel to probe the
Higgs-top Yukawa coupling in case of an IMH.  $t\bar tZ$ production can be
a potential background to $t\bar tH$ if the Higgs mass is close to $m_Z$.
As an alternative to its parent $e^+e^-$ collider, $\gamma\gamma
\rightarrow t\bar tH(Z)$ productions are  compared
with the corresponding productions in the $e^+e^-$ collisions.
\end{abstract}

\newpage
\section{INTRODUCTION}
\label{intro}

Ever since the discoveries of $W$\cite{W} and $Z$\cite{Z} bosons, the
Standard Model (SM) has been tested to high accuracy \cite{lang}.
However, the spontaneous symmetry-breaking is not yet well understood,
nor is there any experimental evidence to favour any particular
symmetry-breaking model.  The simplest model is the Minimal SM
with a single neutral scalar Higgs boson \cite{book} to activate the
Higgs mechanism \cite{higgs}.  One of the major
goal of the next generation $pp$, $e^+e^-$, $ep$ and
the newly discussed $\gamma\gamma$ colliders is to look for the Higgs
boson \cite{hunter}.
For  heavy Higgs ($m_H \agt 2m_Z$) the $W^+W^-$\cite{cheung-ww,cheung-ee} and
$ZZ$\cite{cheung-ee,cheung-zz} decay modes have been shown to be
viable channels for detection in future $e^+e^-$ and $pp$ colliders.
More troublesome  is the intermediate mass Higgs,
one must look for $H\rightarrow \gamma\gamma$\cite{gamma},
$WW^*$ and $ZZ^*$\cite{gour}, and $b \bar b$ or $\tau\tau$\cite{bbbar} modes,
and the feasibility depends sensitively on the resolution of the
detectors\cite{dete}.

On the other hand, the top-quark is very likely to exist because there
must be an $SU(2)_L$ isospin partner to the $b$-quark.
In the minimal SM the fermions acquire masses via their $Yukawa$ coupling
to the Higgs.  At tree level the  Higgs-boson  couples to a fermion of
mass $m_f$ with strength
\begin{equation}
\label{g_ffH}
     g_{ffH} = -i \frac{g m_f}{2 m_W}\, ,
\end{equation}
where $g$ is the $SU(2)$ gauge coupling.
The coupling in Eq.~(\ref{g_ffH}) can be directly probed in the decay of the
Higgs boson into a pair of fermions if kinematically allowed.
To measure $g_{ttH}$ directly by this method however we need a Higgs of
mass $> 2 m_t$.  The top-quark mass
is likely in the range 120$\sim$200~GeV, so  we need a
Higgs-boson of mass greater than about 250~GeV to allow the decay into a
$t\bar t$ pair.  Consequently, if the Higgs mass lies within the
intermediate mass range ($m_W<m_H < 2m_Z$), this method cannot be used
to probe the $g_{ttH}$ coupling directly.
It can be probed indirectly  in the decay of $H\rightarrow
\gamma\gamma$ or $gg$ (see {\it e.g.} Refs.~\cite{book,hunter}) or the
fusion of $\gamma\gamma\rightarrow H$\cite{bord} or
$gg\rightarrow H$\cite{gluon}
through an internal  top-quark loop; but it is likely to be
affected by the presence of other heavy particles beyond the SM.
An alternative  direct probe is to use the associated production of a
Higgs-boson  with a $t\bar t$ pair at the $e^+e^-$\cite{djo} and
$pp$\cite{gunion} colliders.
In principle, the same coupling can also be probed  in the production process
 $e^+e^- \rightarrow t\bar tZ$ \cite{hagi}, but the contribution
{}from the Higgs-exchange diagram
is very small relative to the contributions from other diagrams
unless the Higgs mass is above the $t\bar t$ threshold,
so the $e^+e^-\rightarrow t\bar tZ$ production is very
insensitive to the presence of an IMH.
Therefore, in the case of IMH, $t\bar tZ$ production is not a good channel to
probe the Higgs-top coupling, but rather  a potential
background to $t\bar tH$ production, especially if $m_H$ is close to
$m_Z$.

With the new possibility of $\gamma\gamma$ collisions \cite{bord,teln}
at $e^+e^-$
colliders, the production process $\gamma\gamma \rightarrow t \bar t H$
offers another possible direct test of the $g_{ttH}$ coupling in addition
to $t\bar tH$ production in $e^+e^-$
and hadronic collisions.  The $\gamma\gamma$ collisions at  $e^+e^-$
machines can be  realized by shining a low energy (a few $eV$) laser beam at a
very small angle $\alpha_0$, almost head to head, to the incident
electron beam.  By Compton
scattering, there are abundant, hard back-scattered photons in the same
direction as the incident electron, which carry a substantial fraction of
the energy of the incident electron.  Similarly, another laser beam
can be directed onto the positron beam, and the resulting $\gamma$ beams
effectively make a $\gamma\gamma$ collider.   Actually, the second beam
need not be positrons, but could also be electrons.  For further technical
details please see Ref.~\cite{teln}. Another possibility is to use the
beamstrahlung effect \cite{beams} but this method produces photons
mainly in the soft region \cite{teln}, and depends critically on the beam
structure \cite{beams}. For the productions of $t\bar tH$ and $t\bar tZ$
we would need a high energy threshold of the beamstrahlung photons.
Therefore we shall limit our calculations to $\gamma\gamma$ collisions
produced by the laser back-scattering method.

The best signal for $t\bar t H$ production will be due to the dominant
decay modes  $H \rightarrow b\bar b$ and $t \rightarrow bW$, and
therefore the signature is
\begin{equation}
\label{bbbbWW}
\gamma \gamma \rightarrow t\bar t H\rightarrow b\bar b b\bar b WW\, .
\end{equation}
Since these rare events will only be searched for after the discovery
of the Higgs-boson,
backgrounds can be removed by using the constraints due to the  $W$,
$t$ and $H$ masses.  Even so, $t\bar tZ$ production is a potential
background, especially if the Higgs mass is close to the Z mass; {\it
e.g.}, $|m_H -m_Z|$ less than a few GeV.  Tagging the $b$ would be very
helpful, but $b$-tagging efficiency is, so far, quite uncertain.
If $b$-tagging has a high efficiency, then the $t\bar tZ$ background can
be reduced substantially.
There might be kinematic regions where $t\bar tH$ production dominates
$t\bar tZ$ production in $\gamma\gamma$ collisions even though $t\bar tZ$
production is larger than $t\bar tH$ production in $e^+e^-$ collisions.
For example, for $m_t=150$ GeV and  $m_H=100$ GeV,
$\sigma(e^+e^- \rightarrow t \bar tH) \simeq 2$ fb,
and $\sigma(e^+e^- \rightarrow t\bar tZ) \simeq 5$ fb for
$e^+e^-$ collisions at 1~TeV.
However, it was found in Ref.~\cite{halzen} that $t\bar t$
production in $\gamma\gamma$ collisions realized by laser back-scattering
is slightly larger than the direct $e^+e^- \rightarrow t\bar t$
production for $m_t\alt 130$~GeV at $\sqrt{s}=0.5$~TeV;
and at $\sqrt{s}=1$~TeV the production of
$\gamma\gamma \rightarrow t\bar t$ is much larger than
$e^+e^- \rightarrow t\bar t$ for $m_t \sim 100-200$~GeV both with and
without considering the threshold QCD effect.
In the following we will explore how feasible the $\gamma\gamma$ collider
is for $t\bar tH$ production, which will then directly probe the $g_{ttH}$
Yukawa coupling.
In Sec.~\ref{II} we will present the calculation methods,
which include the photon luminosity and subprocess cross sections.
The results are discussed in Sec.~\ref{III}, and in Sec.\ref{IV} the
conclusions are summarized.

\section{CALCULATION METHODS}
\label{II}

\subsection{Photon Luminosity}

Using the laser back-scattering technique on an electron- or
positron-beam abundant numbers of hard photons can be produced
nearly in the same direction as the original beam.  A low energy $\omega_0$
(a few $eV$) laser beam is directed onto the electron beam almost head to
head.  The energy $\omega$ of the scattered photon depends on its angle
$\theta$ with respect to the  incident electron beam and is given by
\begin{equation}
\omega = \frac{E_0 (\frac{\xi}{1+\xi})}{1+(\frac{\theta}{\theta_0})^2} \, ,
\end{equation}
where
\begin{equation}
\theta_0 =\frac{m_e}{E_0} \sqrt{1+\xi}\,, \,  \; \xi =\frac{4E_0
\omega_0}{m_e^2}  \,,
\end{equation}
and $E_0$ is the energy of the incident electron.  Therefore, at $\theta=0$,
$\omega =E_0\xi/(1+\xi)=\omega_{\rm max}$ is the maximum energy of the
back-scattered photon.  The energy spectrum of the back-scattered photon,
shown in Fig.~\ref{spectrum}, is given by \cite{teln}
\begin{equation}
\label{lum}
F_{\gamma /e}(x) = \frac{1}{D(\xi)} \left[ 1-x +\frac{1}{1-x}
-\frac{4x}{\xi(1-x)} + \frac{4x^2}{\xi^2 (1-x)^2} \right] \,,
\end{equation}
where
\begin{equation}
\label{D_xi}
D(\xi) = (1-\frac{4}{\xi} -\frac{8}{\xi^2}) \ln(1+\xi) + \frac{1}{2} +
\frac{8}{\xi} - \frac{1}{2(1+\xi)^2}\,,
\end{equation}
and $x=\omega/E_0$ is the fraction of the energy of the incident
electron carried by the back-scattered photon.   Therefore
\begin{equation}
x_{\rm max}=\frac{\omega_{\rm max}}{E_0} = \frac{\xi}{1+\xi}\,,
\end{equation}
is the maximum fraction of energy carried away by the back-scattered photon.
{}From Eq.~(\ref{lum}) and (\ref{D_xi}) the portion of photons with
energy close to
$\omega_{\rm max}$ grows with $E_0$ and $\omega_0$, and so does
$x_{\rm max}$.  However, we should not choose a large $\omega_0$, or the
back-scattered photon will interact with the incident photon and create
unwanted $e^+e^-$ pairs.  The threshold for  $e^+e^-$ pair creation is
$\omega_{\rm max} \omega_0 > m_e^2$, so we require $\omega_{\rm
max} \omega_0 \alt m_e^2$.  Solving $\omega_{\rm max}\omega_0=m_e^2$, we find
\begin{equation}
\xi = 2(1+\sqrt{2}) \simeq 4.8 \,.
\end{equation}
For the choice $\xi=4.8$ one finds $x_{\rm max}\simeq 0.83$,
$D(\xi)\simeq 1.8$, and
\begin{eqnarray}
\omega_0 & = & \frac{\xi m_e^2}{E_0} \, ,  \nonumber \\
&=&  \left \{    \begin{array}{cl}
                1.25\, {\rm eV} \quad & \mbox{for a 0.5 TeV $e^+e^-$
collider} \\
	   	0.63\, {\rm eV} \quad & \mbox{for a 1 TeV $e^+e^-$\,
collider.}
		  \end{array}
     \right.
\end{eqnarray}
Here we assume that the average polarization of the back-scattered photon is
zero; i.e., an unpolarized $\gamma$-beam.  We also assume that, on average,
the number of the back-scattered photons produced per electron is 1, i.e.,
the conversion coefficient $k$ is equal 1.

\subsection{Subprocesses}

For $\gamma\gamma \rightarrow t\bar tH$
the contributing Feynman diagrams are shown in Fig.~\ref{feynman}, in
which the cross diagrams with the interchange of the two incoming photons are
not shown.  The Higgs can be  radiated from any fermion-line, so each
diagram is proportional to $g_{ttH}$ and  the resulting cross section
will then be proportional to $g_{ttH}^2$.  Consequently, this process
directly probes the Higgs-top coupling.  The amplitudes for the
contributing Feynman diagrams are given in Appendix~\ref{appI}.  The
corresponding Feynman diagrams for $\gamma\gamma \rightarrow t\bar
tZ$ can be derived from those in Fig.~\ref{feynman} by simply replacing the
Higgs by the $Z$.  These Feynman amplitudes are also given in
Appendix~\ref{appI}.   The subprocesses $e^+e^- \rightarrow t\bar tH$
\cite{djo} and $t\bar tZ$ \cite{hagi} have been previously calculated,
and it is not necessary to repeat these formulas here.
We, however, independently did the calculations and
agree with the results in Refs.~\cite{djo} and \cite{hagi}, respectively.

To obtain the total cross sections $\sigma$ we fold in the photon luminosity
with the cross section $\hat \sigma$ for the subprocesses. The resulting
total cross section $\sigma$ is
\begin{equation}
\sigma = \int_{x_{1min}}^{x_{max}}  \int_{x_{2min}}^{x_{max}}
F_{\gamma/e}(x_1) F_{\gamma/e}(x_2) \hat \sigma (\gamma\gamma\rightarrow
t\bar tV\;{\rm at}\; \hat s = x_1 x_2 s)
dx_1 dx_2 \,,
\end{equation}
with the constraints
\begin{eqnarray}
x_1,\,x_2  & \leq &  x_{\rm max} \, , \\
\frac{(2m_t+m_V)^2}{s} &  \leq &  x_1 x_2\,, \quad {\rm where}\;
V  =  H,\, Z\,.
\end{eqnarray}
Throughout the paper, $\sqrt{s}$ always refers to the center-of-mass
energy of the $parent$ $e^+e^-$ collider.

\section{RESULTS \& DISCUSSION}
\label{III}

We have used the energy spectrum of back-scattered photons shown in
Fig.~\ref{spectrum}. With the choice of $\xi=4.8$ a large fraction
($>50 \%$) of the photons  have energies greater than $0.5E_0$ and the
spectrum peaks at the end point $x_{\rm max}\simeq 0.83$.
We shall consider, typically, $m_t >120$ GeV, which satisfies
the CDF  95\% confidence level bound of $\agt 91$ GeV \cite{CDF},
and the range $m_H \sim 60-140$ GeV. The energy threshold
for $t\bar tH$ production will then be  at least 300 GeV.  The corresponding
threshold value for $z=\sqrt{x_1x_2}=\sqrt{\hat s/s}$ is 0.6, 0.3 and 0.15 for
0.5, 1 and 2 TeV $e^+e^-$ colliders, respectively.  At these high values of
$z$, the gluon content inside the ``resolved'' photon is negligible (see
Fig.~1(b) and 1(c) of Ref.~\cite{halzen}).  Therefore, we need only to
consider direct $\gamma\gamma$ collisions.

In Fig.~\ref{XvsE} we show  the total cross sections as a function of the
center-of-mass energy $\sqrt{s}$ of the parent $e^+e^-$ collider for
$m_H=90$~GeV and $m_t=120$ and 150~GeV.  We have ignored beamstrahlung and
bremsstrahlung of the initial states in the calculations of
$e^+e^- \rightarrow t\bar tH,\,t\bar tZ$.  For $e^+e^-$ collisions, both
$t\bar tH$ and $t\bar tZ$ productions reach a maximum between about
$\sqrt{s}=500$~GeV to 750 GeV, and then fall gradually as $\sqrt{s}$
increases further.  There are two main factors for this feature:
one is the phase space factor and another is the $1/s$ factor in
the $s$-channel  $\gamma$ or $Z$ propagator.
 As $\sqrt{s}$ first increases from 500 GeV, the phase space factor
increases; but as $\sqrt{s}$ increases further, the increase in phase
space factor is offset by the $1/s$ decrease of the propagator.
Roughly, $t\bar tZ$ production in $e^+e^-$ collisions is about a
factor of 2 to 5 larger than production of $t\bar tH$.  Consequently, if
$m_H$ is close to $m_Z$ the $t\bar tH$ signal could be difficult to identify
due to the potential $t\bar tZ$ background.

On the other hand, the cross sections for
$\gamma\gamma \rightarrow t\bar tH$ and $t\bar tZ$ start off very
small at $\sqrt{s}=500$~GeV because they are very limited by
the luminosity function $F_{\gamma/e}(x)$ at this energy.
But both increase gradually as $\sqrt{s}$ increases, because a growing
range of $x$ is available and there is no propagator contributing
a factor $1/s$ as in the corresponding case of $e^+e^-$ collisions.
However, $\sigma(\gamma\gamma\rightarrow t\bar tH)$ begins to flatten
 out after $\sqrt{s}=1.5$~TeV.  For both values of $m_t$ $t\bar tH$
production is larger than $t\bar tZ$ production at lower energies. But
$t\bar tZ$ increases above
$t\bar tH$ at about $\sqrt{s}=1(2)$~TeV for $m_t=120(150)$~GeV.
For $m_t=150$~GeV we have a $t\bar tH$ signal larger than the
potential $t\bar tZ$ background for the entire range of $\sqrt{s}$ from
0.5 to 2 TeV.  This is an important advantage of $\gamma\gamma$
collisions over $e^+e^-$ collisions for directly probing the $g_{ttH}$
Yukawa coupling.
For $\sqrt{s}$ from 0.5 TeV to about 1.1
TeV, the $e^+e^-\rightarrow t\bar tH$ cross sections are larger than the
$\gamma\gamma\rightarrow t\bar tH$ cross sections. However, for this
range of $\sqrt{s}$, the potential background from $t\bar tZ$ production
is also greater in $e^+e^-$ collisions.
As $\sqrt{s}$ increases further $\gamma\gamma$ collisions are more
advantageous both because $\sigma(\gamma\gamma\rightarrow t\bar tH)$ is
larger and because there is a smaller $t\bar tZ$ background.

In Fig.~\ref{XvsMH}, we plot the variation of total cross sections with
Higgs mass $m_H$ for the range 60--140~GeV and $m_t=150$~GeV at
$\sqrt{s}=1$ and 2~TeV.  Of expected, $t\bar tH$ production in both
$e^+e^-$ and $\gamma\gamma$ collisions decreases with increasing $m_H$,
simply because less phase space is available.  In contrast
$\gamma\gamma \rightarrow t\bar tZ$ is independent of $m_H$ and the
effect of $m_H$ on
$e^+e^- \rightarrow t\bar tZ$ is negligible since the Higgs-exchange
diagram is insignificant for this range of $m_H$.

In Fig.~\ref{XvsMT} we show the dependence of the total cross sections
on the top-quark mass $m_t$ for $m_H=90$~GeV at $\sqrt{s}=1$ and 2 TeV.
Two factors dominate the main features of these curves:
phase space and $g_{ttH}$ coupling.
The coupling $g_{ttH}$ grows linearly with increasing $m_t$.  Therefore both
$e^+e^-,\,\gamma\gamma \rightarrow t\bar tZ$ decrease as $m_t$
increases simply  because less phase space becomes available, and the
effect is more pronounced at the smaller value of $\sqrt{s}$.
On the other hand, at $\sqrt{s}=1$ and 2~TeV both
$e^+e^-,\,\gamma\gamma \rightarrow t\bar tH$ productions are
enhanced as $m_t$ increases because the increase in the coupling
$g_{ttH}$ is more important than the decreasing phase space.
At $\sqrt{s}=1$~TeV, $\gamma\gamma \rightarrow t\bar tH$ however  begins
to flatten out after about $m_t$=160 GeV.

\section{Conclusion}
\label{IV}
The Next Linear Colliders(NLC) will have center-of-mass energies
{}from 0.5 to 2 TeV.  The Yukawa coupling $g_{ttH}$ can be probed directly
via $t\bar tH$ production, although there is a
potential background from $t\bar tZ$ production if $m_H$ lies
close to $m_Z$.  For $\sqrt{s}$ from 0.5 to 1 TeV,
the $e^+e^-\rightarrow t\bar tH$ cross section (1.5--3~fb) is larger than
$\sigma(\gamma\gamma\rightarrow t\bar tH)$, but so is the potential
background: $\sigma(e^+e^- \rightarrow t\bar tZ) \sim 3$--6~fb.
For $\sqrt{s}\agt 1$~TeV $\gamma\gamma$ collisions provide a better
approach than $e^+e^-$ collisions since the cross section ($\sim
1.5-2$~fb) is larger and there is less potential background from
$t\bar tZ$ production ($\sim 0.5-2.5$~fb).
For a yearly luminosity of 10~${\rm fb}^{-1}$, $m_t=150$~GeV and
$m_H=90$~GeV we have
about 14(22) $t\bar tH$ events and about 6(21) $t\bar tZ$ events in
$\gamma\gamma$ collisions realized by the laser back-scattering method
at 1(2)~TeV $e^+e^-$ colliders.

Added Note: after completing this work, we came across a paper by E.~Boos {\it
et al.}\cite{BOOS}.  On the part of overlapping, our results agree with theirs.

\newpage
\acknowledgements
Special thanks to R.~Oakes for reading the manuscript and giving a lot of
invaluable comments on the work.  Also thanks to E.~Braaten and D.~Chang for
reading the manuscript.
This work was supported by the U.~S. Department of Energy, Division of
High Energy Physics, under Grant DE-FG02-91-ER40684.
\appendix{}
\label{appI}

This appendix gives the formulas for the Feynman amplitudes of the
subprocesses $\gamma\gamma \rightarrow t\bar tH$ and $\gamma\gamma
\rightarrow t\bar tZ$ (see Fig.~\ref{feynman}).
Defining some notations as follows :
\begin{eqnarray}
D^t(k) & =  &\frac{1}{k^2 - m_t^2}\,,\\
g^V(t) &=& g^V_v (t) + g^V_a(t) \gamma^5 \,, \quad {\rm where}\;
V=\gamma,\,Z \,.
\end{eqnarray}
The couplings are given by
\begin{equation}
\begin{array}{rcl}
g^Z_v(f) & = & g_Z ( \frac{T_{3f}}{2} - Q_f x_W ) \,, \\
g^Z_a(f) & = & - g_Z \frac{T_{3f}}{2} \,, \\
g^\gamma_v (f) &= & g \sin\theta_W Q_f\,,\\
g^\gamma_a (f) &=& 0 \,,
\end{array}
\end{equation}
where $g_Z = g/ \cos\theta_W$ and $f$ is a fermion.

The amplitudes for
the $\gamma(p_1)\gamma(p_2) \rightarrow t(q_1)\bar t(q_2)H(k_1)$ are given by
\begin{eqnarray}
{\cal M}^{(a)} &=& - \frac{gm_t}{2m_W} D^t(q_1+k_1) D^t(p_2-q_2) \nonumber \\
&& \quad \times \bar u(q_1) (\overlay{/}{q}_1 + \overlay{/}{k}_1 +m_t)
\overlay{/}\epsilon (p_1) g^\gamma(t) (\overlay{/}{p}_2-\overlay{/}{q}_2
+ m_t) \overlay{/}\epsilon(p_2) g^\gamma(t) v(q_2)\,, \\
{\cal M}^{(b)} &=& - \frac{gm_t}{2m_W} D^t(q_1-p_1) D^t(p_2-q_2) \nonumber \\
&& \quad \times \bar u(q_1) \overlay{/}\epsilon(p_1) g^\gamma(t)
(\overlay{/}{q}_1 - \overlay{/}{p}_1 +m_t)
(\overlay{/}{p}_2-\overlay{/}{q}_2 + m_t)
\overlay{/}\epsilon(p_2) g^\gamma(t) v(q_2)\,, \\
{\cal M}^{(c)} &=& - \frac{gm_t}{2m_W} D^t(q_1-p_1) D^t(q_2+k_1)\nonumber \\
&& \quad \times \bar u(q_1) \overlay{/}\epsilon(p_1) g^\gamma(t)
(\overlay{/}{q}_1 - \overlay{/}{p}_1 +m_t)
\overlay{/}\epsilon(p_2) g^\gamma(t)
(-\overlay{/}{q}_2-\overlay{/}{k}_1 + m_t) v(q_2) \,,
\end{eqnarray}
with the addition of the terms by interchanging $\gamma(p_1)
\leftrightarrow \gamma(p_2)$.
The Feynman amplitudes for $\gamma(p_1) \gamma(p_2) \rightarrow t(q_1)
\bar t(q_2) Z(k_1)$ are given by
\begin{eqnarray}
{\cal M}^{(a)} &=& -  D^t(q_1+k_1) D^t(p_2-q_2) \nonumber \\
&& \quad \times \bar u(q_1) \overlay{/}\epsilon(k_1) g^Z(t)
(\overlay{/}{q}_1 + \overlay{/}{k}_1 +m_t)
\overlay{/}\epsilon (p_1) g^\gamma(t)
(\overlay{/}{p}_2-\overlay{/}{q}_2 + m_t)
\overlay{/}\epsilon(p_2) g^\gamma(t) v(q_2)\,, \\
{\cal M}^{(b)} &=& -  D^t(q_1-p_1) D^t(p_2-q_2) \nonumber \\
&& \quad \times \bar u(q_1) \overlay{/}\epsilon (p_1) g^\gamma(t)
(\overlay{/}{q}_1 - \overlay{/}{p}_1 +m_t)
\overlay{/}\epsilon(k_1) g^Z(t)
(\overlay{/}{p}_2-\overlay{/}{q}_2 + m_t)
\overlay{/}\epsilon(p_2) g^\gamma(t) v(q_2)\,, \\
{\cal M}^{(c)} &=& -  D^t(q_1-p_1) D^t(q_2+k_1) \nonumber \\
&& \quad \times \bar u(q_1) \overlay{/}\epsilon (p_1) g^\gamma(t)
(\overlay{/}{q}_1 - \overlay{/}{p}_1 +m_t)
\overlay{/}\epsilon(p_2) g^\gamma(t)
(-\overlay{/}{q}_2-\overlay{/}{k}_1 + m_t)
\overlay{/}\epsilon(k_1) g^Z(t)v(q_2)\,,
\end{eqnarray}
with the addition of the terms by interchanging $\gamma(p_1)
\leftrightarrow \gamma(p_2)$.

\newpage
\figure{\label{spectrum}
The energy spectrum $F_{\gamma/e}(x)$ of the back-scattered photon versus
the energy fraction $x$ of the incident electron being carried away by
the back-scattered photon.}

\figure{\label{feynman}
Contributing Feynman diagrams for the process $\gamma\gamma \rightarrow
t \bar t H$.  Cross diagrams by interchanging the two incoming photons
are not shown.}

\figure{\label{XvsE}
Total cross sections versus center-of-mass energy of the parent $e^+e^-$
collider, for $m_H=90$ GeV and $m_t=$ (a) 120, (b) 150~GeV.
The subprocesses $\gamma\gamma \rightarrow t\bar tH$ (solid), $t\bar tZ$
(dashed); $e^+e^- \rightarrow t\bar tH$ (dotted), $t\bar tZ$
(dash-dotted) are shown.}

\figure{\label{XvsMH}
Total cross sections versus the mass $m_H$ of Higgs-boson for $m_t=150$~GeV,
and  $\sqrt{s}$= (a) 1 TeV, and (b) 2 TeV.
The subprocesses $\gamma\gamma \rightarrow t\bar tH$ (solid), $t\bar tZ$
(dashed); $e^+e^- \rightarrow t\bar tH$ (dotted), $t\bar tZ$
(dash-dotted) are shown.}

\figure{\label{XvsMT}
Total cross sections versus the top-quark mass $m_t$ for $m_H=90$~GeV and
$\sqrt{s}=$ (a) 1 TeV, and (b) 2 TeV.
The subprocesses $\gamma\gamma \rightarrow t\bar tH$ (solid), $t\bar tZ$
(dashed); $e^+e^- \rightarrow t\bar tH$ (dotted), $t\bar tZ$
(dash-dotted) are shown.}

\end{document}